\documentclass[%
 aip,
 jmp,%
 amsmath,amssymb,
 reprint,%
]{revtex4-1}

\usepackage{graphicx}
\usepackage{dcolumn}
\usepackage{bm}
\usepackage[caption = false]{subfig}

\begin{document}


\title{ Wavelength Dependent Nonlinear Spectroscopic Study of Third Harmonic Generation  Probed by Rotational Maker Fringes Method}

\author{Calford Odhiambo Otieno}
 \affiliation{1. Department of Physics,Kisii University P.O Box 408-40200,Kisii Kenya.}

 \email{cotieno@kisiiuniversity.ac.ke}

\date{\today}

\begin{abstract}
Efficient third-order nonlinearities of the Zinc Oxide and Al-doped Zinc Oxide were studied by Third Harmonic Generation (Third Harmonic Generation) Maker fringes to establish the effect Aluminum  of Aluminum doping (Al-doping) on the cubic nonlinearities. The addition of the Al-dopant to the Zinc Oxide crystal structure results in changes that affected the optical and nonlinear characteristics. Presented results indicate that the magnitude of $\chi^{(3)}$ was enhanced at single experimental wavelengths, however, across the broadband experimental spectrum the effect of Al-doping remained relatively constant. Observed enhancement of third-order nonlinearity was purely from the bound electronic response and the observation is attributed to increased charge carriers and spontaneous polarization in the Zinc Oxide and Al-doped Zinc Oxide crystal structure.
%
\end{abstract}
\pdfoutput=1
\keywords{Nonlinear Optics, Harmonic Generation, Maker Fringes, Zinc Oxide.}
\maketitle
\section{\label{sec:level1}Introduction}
Zinc Oxide (Zinc Oxide) and Al-doped Zinc Oxide (Al-doped Zinc Oxide) are very promising materials for semiconductor device applications\cite{jagadish2011zinc,fortunato2009zinc}. The material has a direct and wide bandgap in the near-ultraviolet (near-UV) spectral region ($E_{g}$ = 3.3 eV)\cite{srikant1998optical} making it ideal for various applications e.g light-emitting diodes(LED's). Exciton binding energy in Zinc Oxide and Al-dopes Zinc Oxide is in the order of $~$60 meV, this large excitonic energy allows for emission to occur even at room tempreature.\cite{jagadish2011zinc,shan2005transparent,reynolds1999valence}. Under ambient conditions, Zinc Oxide and Al-doped Zinc Oxide typically crystallize in the wurtzite structure and are available as large bulk single crystals\cite{yoshioka2005zinc,chen1998plasma}. Al-doped Zinc Oxide  material properties have been extensively studied\cite{srikant1998optical,chen1998plasma,pearton2005recent}.  From the existing materials Nonlinear optical (Nonlinear Optical) properties of Zinc Oxide have been well researched; however, existing results are mostly available at single wavelengths or limited ranges.  However, the relevant Nonlinear Optical dispersions, i.e., wavelength-dependences are not fully studied especially at longer wavelengths, far below the bandgap because doping can induce a drastic change in the
Nonlinear Optical responses at this broad range of wavelengths via doping-induced subgap-state contributions. Moreover, there are only a few reports on the Nonlinear Optical properties of Al-doped at the single wavelength of 1064 nm. In this paper, we explore the impact of Al doping on the third-order Nonlinear Optical properties of Al-doped prepared into thin films by atomic layer deposition (ALD) and systematically study whether Al doping can enhance Nonlinear Optical responses as a function of doping level. We explain our results by considering the existing theories and making adequate references to the available literature.  
\section{\label{sec:level2}Sample Preparation}
In this section we discuss the instrumentation, sample growth procedure, and the requirements for the atomic layer deposition (ALD) of Zinc Oxide and Al-doped thin films. A commercial ALD system from Sundew Technologies was used. This system employs sequential self-limiting surface reaction between the precursors to achieve atomic layer controlled conformal thin film growth, full details on samples preparation are reported elsewhere\cite{dhakal2012transmittance,}.

\section{\label{sec:level2}Third-Order Harmonic Nonlinear Optical Theory}

In this section, we briefly present the  basic theoretical description of Third Harmonic Generation Nonlinear theory and we follow from\cite{Boyd,meredith1982second}. The Macroscopic polarization $P$ is the number of dipole moments per unit volume.  In a linear regime, $P$ is simply related to the electric field $E $ of the incident light by 
\begin{equation}
P \left( t \right) = \epsilon _{0} \chi ^{ \left( 1 \right) }E \left( t \right),
\end{equation} 
where  $  \chi ^{ \left( 1 \right) } $
is the linear susceptibility and  $ \epsilon _{0} $ is the permittivity of free space. To fully describe the nonlinear polarization effect, Eq. (1) can be expressed as a power series in $E$ and given by  
\begin{equation}
 P \left( t \right) = \epsilon _{o} \left[  \chi ^{ \left( 1 \right) }E \left( t \right) + \chi ^{ \left( 2 \right) }E^{2} \left( t \right) + \chi ^{ \left( 3 \right) }E^{3} \left( t \right) +… \right],  
\end{equation} 
where the higher-order susceptibilities $ \chi ^{ \left( n \right) } $ are tensors in general that relate the vector nature of $P$ and $E$ assuming that the nonlinear medium is lossless and dispersionless. 
The second and the third terms in Equation (2) are the second-and third-order nonlinear polarization, respectively. The second-order susceptibility $\chi ^{ \left( 2 \right) } $ can only occur in a material that has no inversion symmetry (noncentrosymmetric), whereas the third-order susceptibility $\chi^{\left(3\right)}$ can occur in any material including amorphous materials. The Nonlinear Optical response is described by a nonlinear wave equation that has time-varying nonlinear polarization acting as a source to the newly generated electric fields in the medium, the wave equation can be readily obtained from Maxwell's equations as 
\begin{equation}
\triangledown ^{2}E-\frac{n^{2}}{c^{2}}\frac{ \partial ^{2}E}{ \partial t^{2}}=\frac{1}{ \epsilon _{o }c^{2}}\frac{ \partial ^{2}P^{NL}}{ \partial t^{2}} 
\end{equation} 
where macroscopic polarization $P$ has both the linear and the nonlinear terms that are written as $ P=P^{ \left(1\right)}+P^{NL} $, where $ P^{\left(1\right)}$ is linear polarization and  $ P^{NL} $is nonlinear polarization. Third harmonic generation (Third Harmonic Generation) which is the focus of this study can be represented by the third term in Equation 2. For an electric field of the form 
\begin{equation}
 E \left( t \right) =Ecos \omega t, 
\end{equation}
incident on a material, the third-order contribution is given by
\begin{equation}
P^{ \left( 3 \right) } \left( t \right) =\frac{1}{4} \epsilon _{o } \chi ^{ \left( 3 \right) }E^{3}cos3 \omega t+\frac{3}{4} \epsilon _{o } \chi ^{ \left( 3 \right) }E^{3}cos \omega t, 
\end{equation}
where the first term describes the response at 3$ \omega$ which is responsible for Third Harmonic Generation. The second term is the so-called optical Kerr effect. It is important to note that Third Harmonic Generation arises from the interaction of three photons of the same frequency with a Nonlinear Optical medium, resulting in the generation of a single photon with thrice the frequency\cite{Boyd,meredith1982second}. To maximize the conversion efficiency, it is necessary to achieve the phase-matching condition. Third Harmonic Generation phase-matching in transparent materials is however difficult to achieve because of a large index match between the fundamental and the Third Harmonic Generation beams, and as a result, the Third Harmonic Generation study is used as a characterization tool. The wurtzite Zinc Oxide and Al-doped Zinc Oxide structures in this study had a preferential growth orientation along the (0001) and the optical axis is therefore along the \textit{c}-axis thereby presenting only two nonvanishing components of $\chi _{ijkl}^{ \left( 3 \right)}$ namely; $\chi _{zzzz}^{ \left( 3 \right) } $and $\chi_{xxxx}^{ \left( 3 \right) } $. With the known preferential growth direction of the samples, each nonvanishing component  of $\chi_{ijkl}^{\left(3 \right)} $ can be characterized with appropriate polarization choice\cite{zappettini2004wavelength,larciprete2006characterization}. In this study  only the $ \chi _{zzzz}^{ \left( 3 \right) } $ was characterized with $ p^{\omega} $and $ p^{3\omega}$ polarization. For the film samples the absolute value of  $  \chi ^{ \left( 3 \right) }$ can be estimated from the relation\cite{larciprete2006characterization}   
\begin{equation}
I_{3 \omega }=\frac{2304 \pi ^{6}}{A} \left( t_{af}^{ \omega } \right) ^{6} \left( t_{fa}^{3 \omega } \right) ^{2} \frac{sin^{2} \Psi }{ \left( n_{ \omega }^{2}-n_{3 \omega }^{2} \right) ^{2}} \vert  \chi ^{ \left( 3 \right) } \vert ^{2}I_{ \omega }^{3} ,
\end{equation}
where $ t_{ij}^{k} $ are the field transmission coefficients at the input and output surface of the sample. $ n_{\omega } $ and $ n_{3 \omega } $ are the refractive indices at the fundamental frequency and third-harmonic frequency, respectively. $ I_{ \omega } $  is the input fundamental beam,  A is the beam area, and $ I_{3 \omega } $ is the Third Harmonic Generation intensity. In Equation 4.1 $\Psi $ is the phase factor and is defined as  
\begin{equation}
\Psi _{Third Harmonic Generation}=\frac{3 \pi L}{ \lambda _{ \omega }} \left( n_{ \omega }cos \theta _{ \omega }-n_{3 \omega }cos \theta _{3 \omega } \right),
\end{equation} 
where $L$ is the interaction length, i.e., the sample thickness, $\lambda_{ \omega } $ is the wavelength of the input beam.$\theta_{\omega} $and $\theta_{3 \omega }$ are propagation angles at fundamental and third harmonic beam. The Third Harmonic Generation intensity depends on the phase-matching factor
\begin{equation}
\Delta k=\frac{6 \pi }{ \lambda } \left( n_{3 \omega }-n_{ \omega } \right) , 
\end{equation}
Considering the thickness of these samples (L= 250 nm),  the Spectral Maker fringes effect is not expected within the experimental wavelength range.  Figure 4.1 shows the absence of oscillations  for L= $\sim$ 0.25 µm thin films used in the experiment. The advantage of characterizing nonlinearity by measuring Third Harmonic Generation is that it only accounts for the fast Nonlinear Optical response. Due to this other contributions to Nonlinear Optical response such as the thermal effects, orientation, and vibration effects are excluded and cannot affect our results\cite{castaneda2006structural,zappettini2004wavelength}.

\begin{figure}[h]
\centering
\includegraphics[width=\linewidth]{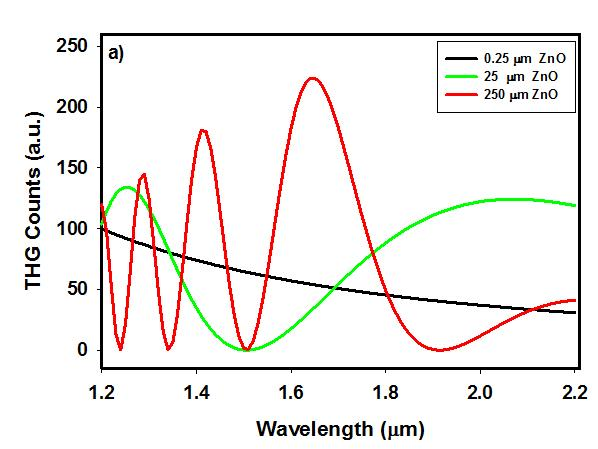}
\caption{Spectral maker fringe for Zinc Oxide, the sample thickness is varied for comparison. There are no oscillations for the L=.25 µm (black line) over the experimental range.}
\label{fig:my_label}
\end{figure}

\section{\label{sec:level3}Experimental Details}
The Third Harmonic Generation properties of the Zinc Oxide and Al-doped Zinc Oxide films were studied for the wavelength ranging from 1.3 µm and 2.1 µm by the Maker fringe. The excitation wavelengths were from a tunable Optical Parametric Oscillator.  The polarization of the fundamental beam was adjusted using a polarizer placed before the sample. Using a collection lens the Third Harmonic Generation signal was directed to a fiber optic bundle connected to the CCD camera. The polarization of the Third Harmonic Generation signal was adjusted using the analyzer after the sample.
The measurements were done in transmission geometry with the sample placed on a rotation stage to allow for continuous angular variation. The rotation axis of the sample was centered on the beam and perpendicular to it to generate symmetric fringe patterns\cite{jerphagnon1970maker,maker1962effects}. Since the glass substrate, as well as other optical components can generate Third Harmonic Generation, careful measurements, and data processing were required to eliminate or minimize these background Third Harmonic Generation signals. First, we used a short-pass filter ($\sim$ 50 $\%$ transmittance in the visible region) just before the fiber-optic cable to suppress the remnant fundamental beam that can cause additional Third Harmonic Generation at the fiber input. Second, we conducted Third Harmonic Generation Z-scan to locate an optimized sample position to maximize the Third Harmonic Generation counts from Zinc Oxide with minimal Third Harmonic Generation from the glass substrate, where the latter was determined by another Third Harmonic Generation Z-scan on a bare glass substrate with the same thickness.
Finally, background Third Harmonic Generation signals from the substrate was measured at each wavelength and subtracted in the Third Harmonic Generation data accordingly to single out Third Harmonic Generation counts from Zinc Oxide only.  A similar procedure was employed for Third Harmonic Generation measurements of Al-doped Zinc Oxide samples.
\section{Results and discussions}
Zinc Oxide and Al-doped exhibited strong angular dependent Third Harmonic Generation Maker fringe when the fundamental wavelength was varied between 1.3 µm and 2.1 µm.  The corresponding Third Harmonic Generation wavelengths ranged from .433 µm to .7 µm. Representative graphs for experimental Third Harmonic Generation Maker fringes for the Zinc Oxide and Al-doped Zinc Oxide samples at $\lambda$ =1.907 µm is shown in Figure 4.2(a-d). The experimental data were fit according to Equations 4.1 and 4.2.
Experimental $\chi^{(3)}$ values were determined using fused silica as a reference, the $\chi^{(3)}$ of fused silica is  $\sim $ 1.99$ \times 10^{-22} m^{2}/V^{2} $
at  $\lambda$ =1.907µm\cite{gubler2000optical,bosshard2000non}.Fused silica was chosen as a reference material since it is a well-studied  material, and its third-order nonlinearity has been determined with high accuracy.

\begin{figure}[h]
\centering
\includegraphics[width=\linewidth]{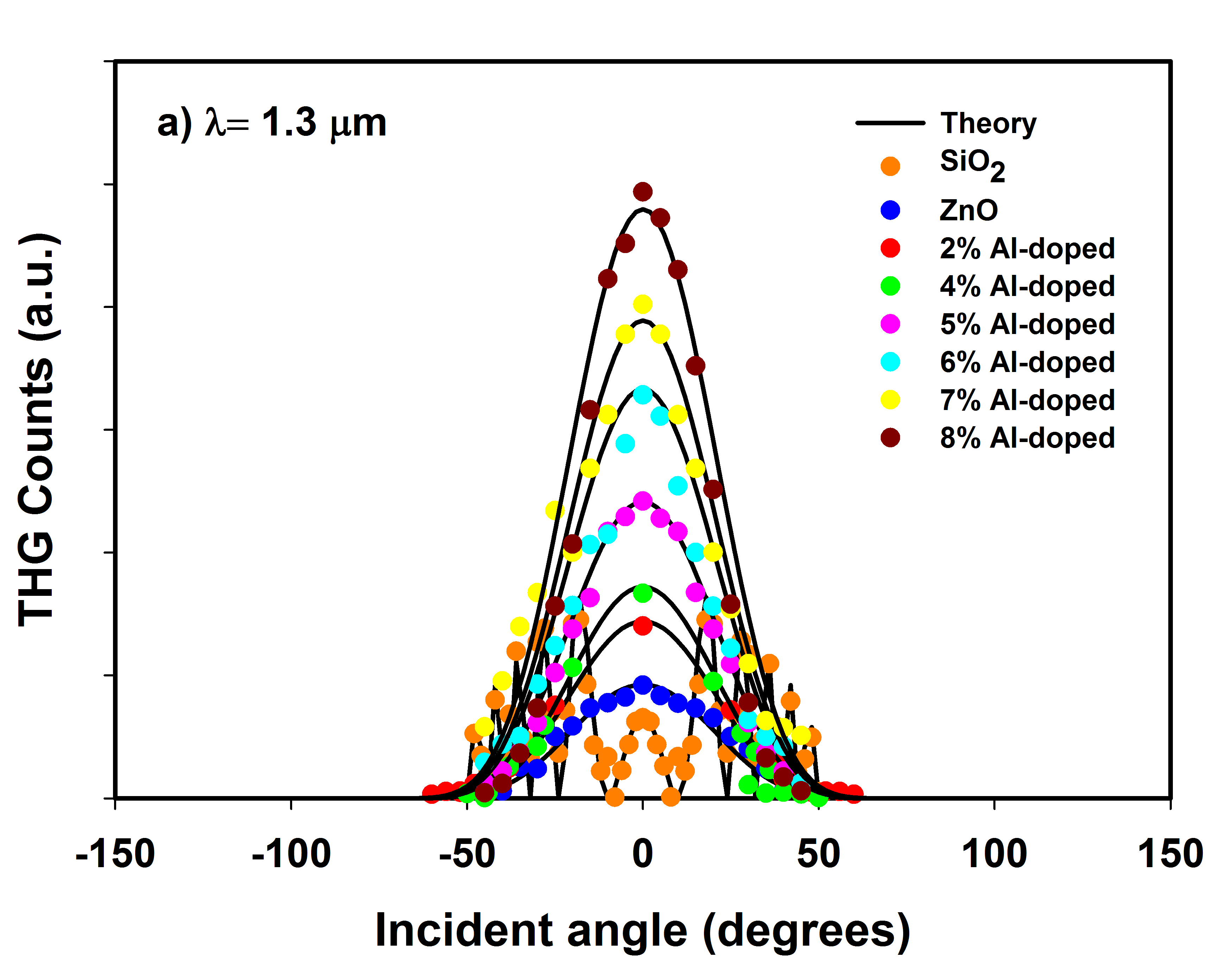}
\caption{Spectral maker fringe for Zinc Oxide, the sample thickness is varied for comparison. There are no oscillations for the L=.25 µm (black line) over the experimental range.}
\label{fig:my_label}
\end{figure}
\begin{figure}[h]
\centering
\includegraphics[width=\linewidth]{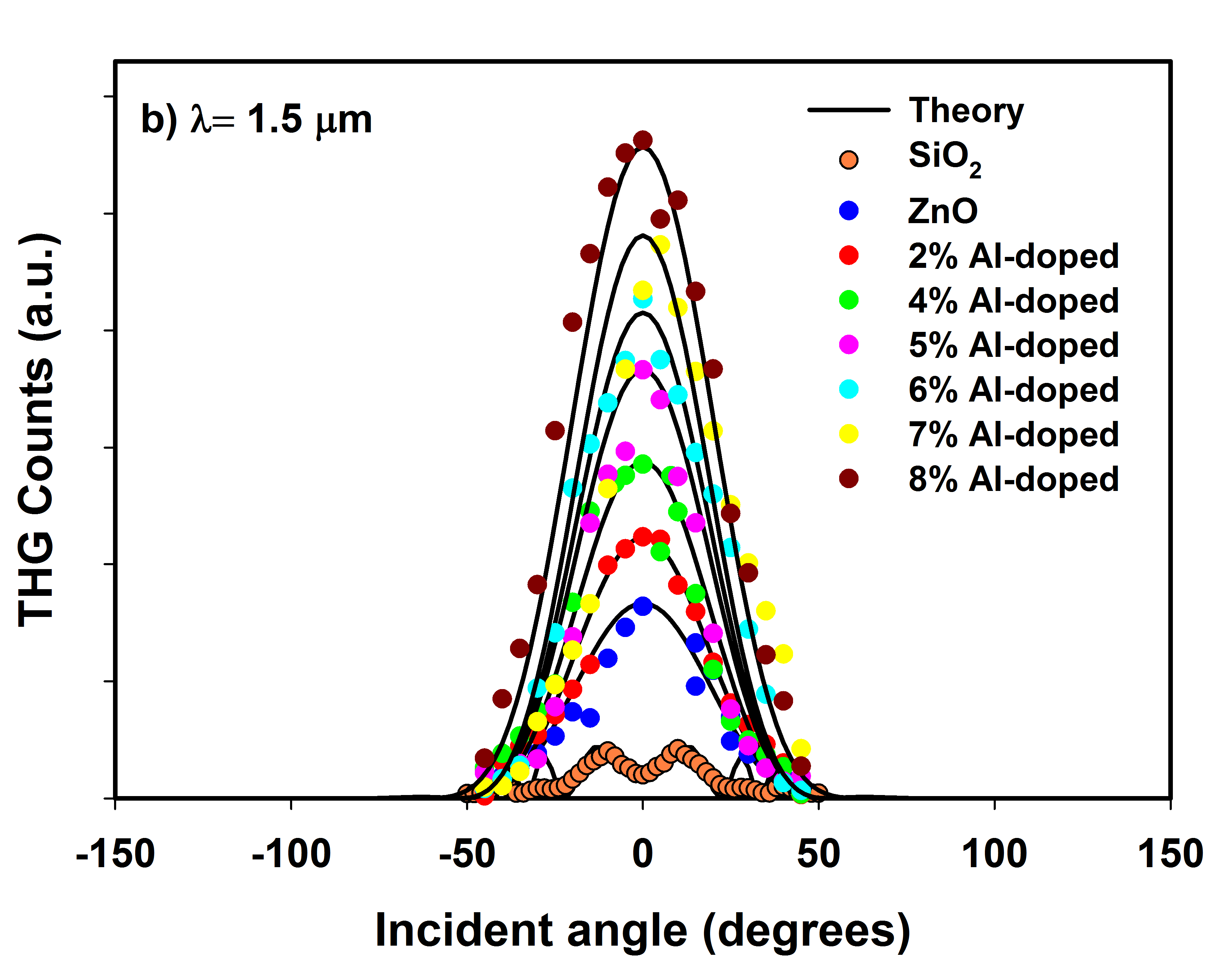}
\caption{Spectral maker fringe for Zinc Oxide, the sample thickness is varied for comparison. There are no oscillations for the L=.25 µm (black line) over the experimental range.}
\label{fig:my_label}
\end{figure}
\begin{figure}[h]
\centering
\includegraphics[width=\linewidth]{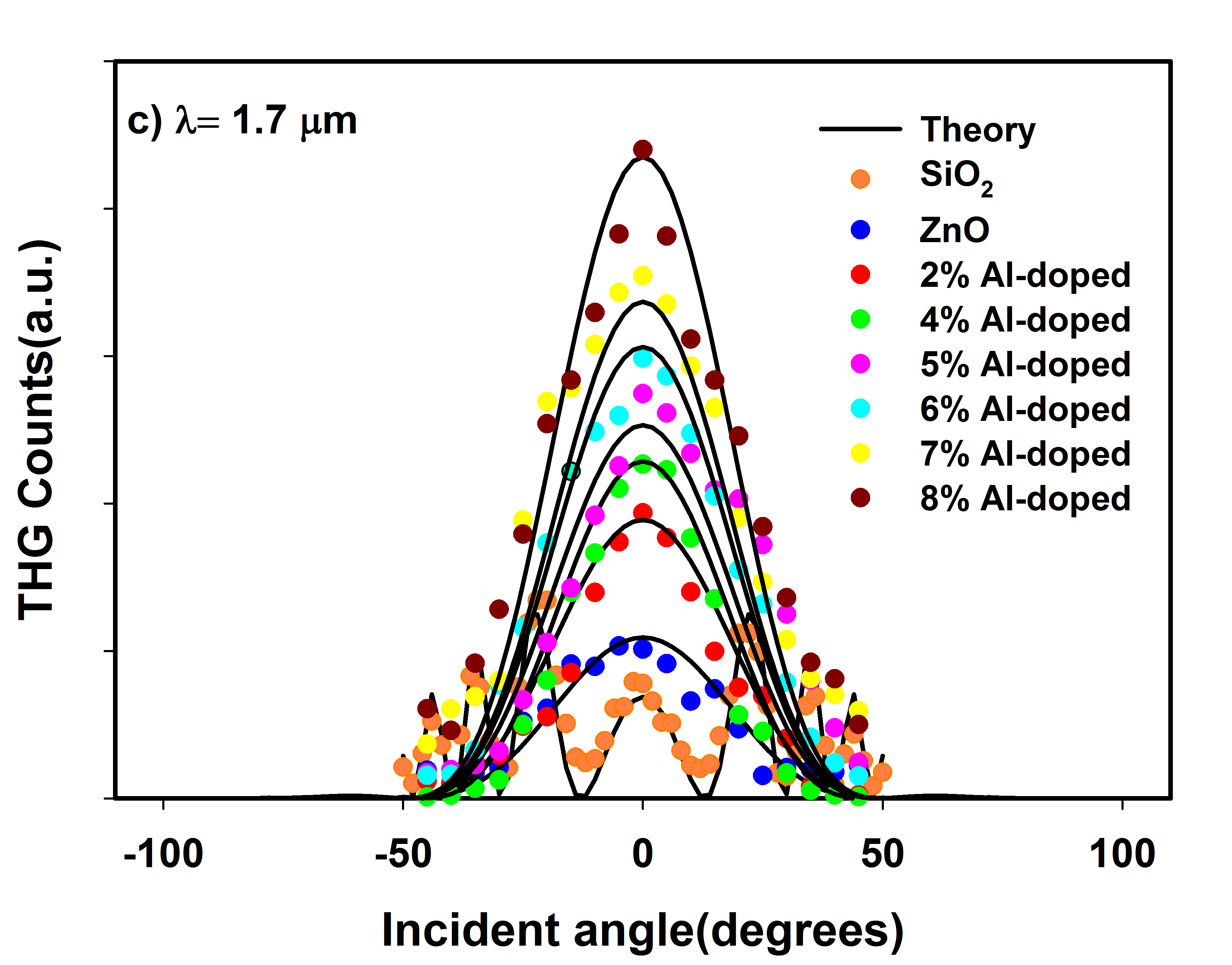}
\caption{Spectral maker fringe for Zinc Oxide, the sample thickness is varied for comparison. There are no oscillations for the L=.25 µm (black line) over the experimental range.}
\label{fig:my_label}
\end{figure}
\begin{figure}[h]
\centering
\includegraphics[width=\linewidth]{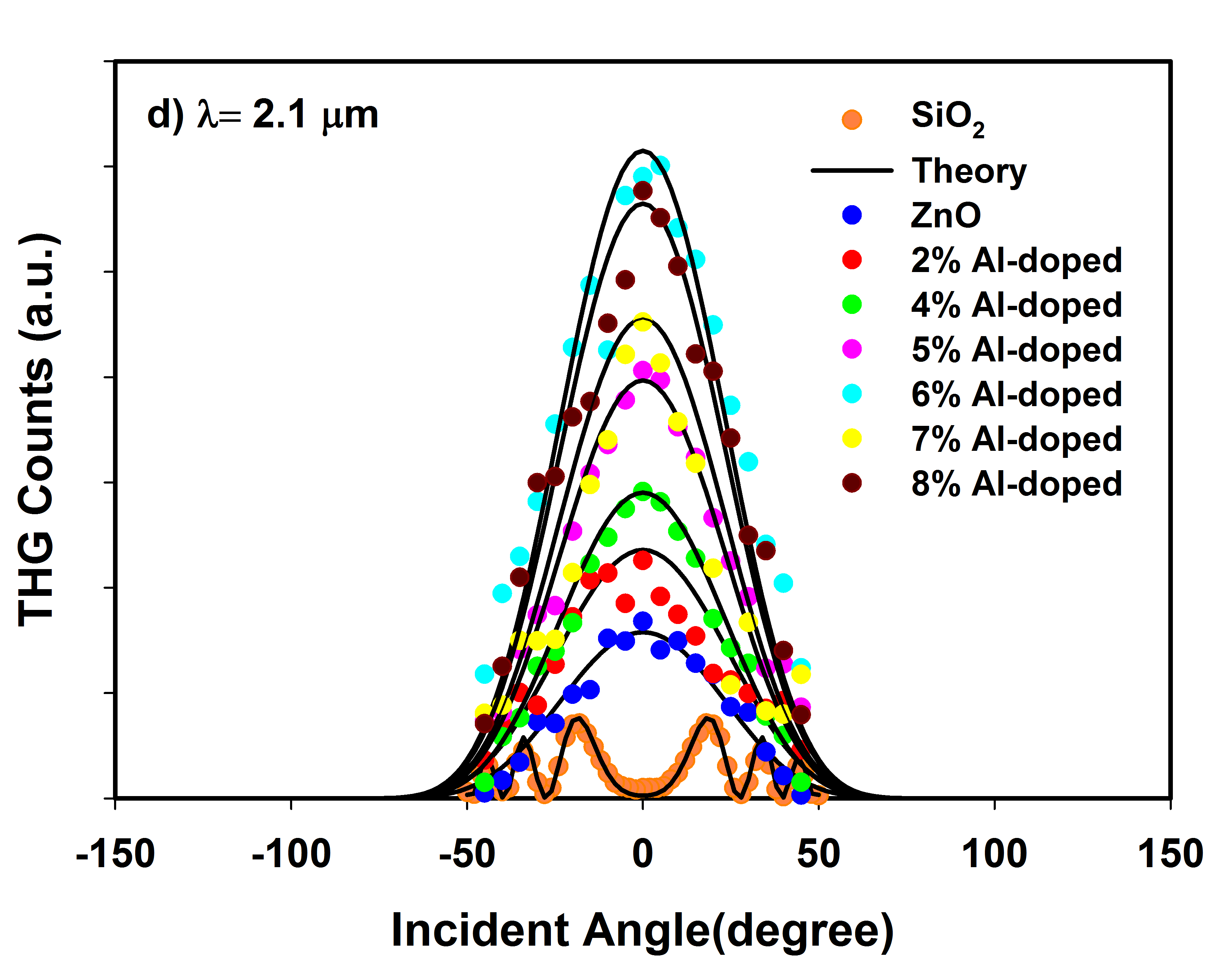}
\caption{Spectral maker fringe for Zinc Oxide, the sample thickness is varied for comparison. There are no oscillations for the L=.25 µm (black line) over the experimental range.}
\label{fig:my_label}
\end{figure}

Since fused silica is centrosymmetric using it as a reference improves the accuracy of our results given its Third Harmonic Generation cannot be affected by any cascade effect. The results indicate that Third Harmonic Generation is dependent on the concentration of the dopant atoms $Al^{3+}$ at specific experimental wavelengths but largely remained constant across the experimental spectrum. This observation is similar to the one reported by Sofiani \textit{et al.} on Al-doped Zinc Oxide grown by spray pyralysis\cite{sofiani2007third,kulyk2007second} Abed et al observed an enhanced Third Harmonic Generation in Nickel doped Zinc Oxide (Ni: Zinc Oxide) as compared to undoped Zinc Oxide\cite{abed2011non}. 
The enhancement of the Third Harmonic Generation signal can be explained by considering the effects of Al$^{3+}$ dopant atoms as follows. Increased concentration of Al$^{3+}$ in Zinc Oxide crystal structures results in two effects; 1) Since $Al^{3+}$ is a donor atom more of it results in increased carrier density (more electrons).  2) Due to the difference in the ionic sizes of Al$^{3+}$ and Zn$^{3+}$, a higher concentration of Al-dopant, the Al$^{3+}$ occupies  interstitial and grain boundary. This overall results in different in-plane orientation unit cells causing the bonds in the neighboring cells to move from the mean position due to Coulomb interaction, this favors dangling bonds at the interface of different unit cells overall resulting in increased change carrier.\cite{liu2004third} 
\begin{figure}[h]
\centering
\includegraphics[width=\linewidth]{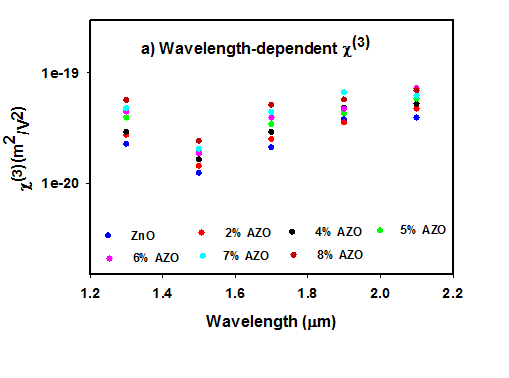}
\caption{Spectral maker fringe for Zinc Oxide, the sample thickness is varied for comparison. There are no oscillations for the L=.25 µm (black line) over the experimental range.}
\label{fig:my_label}
\end{figure}

The enhancement in Third Harmonic Generation observed is thus attributed to the higher carrier densities\cite{liu2004third,neumann2004second,khaled2000generation}. The input field will induce perturbation change on the $Al^{3+}$  causing nonlinear polarization in the media\cite{quimby2006photonics,he1999physics}. A similar argument has also been advanced by Kulyk et al. in thin microcrystalline films of Zinc Oxide\cite{kulyk2009linear,kulyk2007second}.  
The results for broadband dispersion of $  \chi _{zzzz}^{ \left( 3 \right) } $ are presented in Figure 4.3. The spectral dispersion did not show any significant variation over the experimental range, after all the wavelength was varied far from the band gap so no resonance effect was expected. 
The efficient broadband Third Harmonic Generation with suppressed SHG has earlier been observed in Zinc Oxide nanorods prepared by hydrothermal deposition\cite{jang2012strong}. The authors attributed the Third Harmonic Generation response to be bound by the electronic response and suppressed SHG to the micron-size film thickness. For further comparison, we looked at the results of  Zappettini \textit{et. al.}\cite{zappettini2004wavelength} who studied the dispersion of $  \chi _{zzzz}^{ \left( 3 \right) } $ and $  \chi _{xxxx}^{ \left( 3 \right) } $ of Zinc Oxide single crystals in the transparency wavelength range 0.344- 0.540 µm, and presented results showing significantly high values  attributed to exciton enhancement.
In this section, we present a comparative study of the reported  $\chi^{(3)}$ of Zinc Oxide and X: Zinc Oxide (X: Al, Ce, Ga, Sn, F, N, Ag, Cu) at $\lambda$ =1.604 µm with our study. The $\chi^{(3)}$ reported in the literature, and  can be seen from Figure 4.4 the $\chi^{(3)}$ values were found to depend on several factors including; 1) concentration  of the dopant atoms used, 2) thickness of the sample, 3) sample preparation techniques and growth kinetics 4) Excitation source and the wavelength of study. Results of several authors who have studied (Zinc Oxide and Al-doped Zinc Oxide) similar to ours and reported varied results, a few discussed here include; Sofiani \textit{et al.} studied the dependence of $\chi^{(3)}$ on different types of dopants(Cu, Er, Sn, and Al)\cite{sofiani2007third} reported the highest $\chi^{(3)}$ from  Sn-doped Zinc Oxide sample  $  \sim  $ 3.2$\times 10^{-19}m^{2}/V^{2} $(see Figure 4.4).  Castaneda studied F-doped Zinc Oxide and obtained  $\chi^{(3)}$ $ \sim  $  1.2$  \times 10^{-22}m^{2}/V^{2} $ more comparison are in the Figure 4.4 below for $\chi^{(3)}$ studied at 1064 nm. The units are all in MKS and converted according to to\cite{boyd2014third}. 
\subsection{Conclusions}
Efficient third-order nonlinearities of Zinc Oxide and Al-doped Zinc Oxide were studied by Third Harmonic Generation Maker fringes
to establish the effect of doping on the cubic nonlinearities. The addition of the Al-dopant
to the Zinc Oxide crystal structure results in changes that affect the optical and nonlinear characteristics of Zinc Oxide. Our results indicate a slight enhancement of the Third Harmonic Generation at experimental
wavelengths, however, we noted that Third Harmonic Generation remains constant over the experimental range.
We discuss the possible causes of these observations based on the available models and we
concluded that the slight enhancement is due to the increased charge carries with increased
Al-doping. We provided unique results as a function of wavelength and doping. Further study is proposed for different dopants in order to present an opportunity and guideline on the choice of the best dopant for the highest Third Harmonic Generation conversion.
The author declares no conflict of interest and no funding for this research.
\bibliographystyle{apsrev}
\bibliography{references}

\end{document}